# Impact of seed density on continuous ultrathin nanodiamond film formation: an analytical approach

M. Tomellini[*], R. Polini[*]

Dipartimento di Scienze e Tecnologie Chimiche, Università di Roma "Tor Vergata", Via della Ricerca Scientifica 1, Rome, 00133, Italy

**Abstract**

An analytical mean field approach for describing the time evolution of film growth by seeding has been developed. The modeling deals with the generic case of anisotropic growth with different growth rates, respectively on- and normal to- the substrate plane. The finite size of the seeds is considered by including spatial correlation effects among seeds through hard-core interactions. The approach, based on probability theory, provides solution in closed form for mean film thickness as a function of substrate coverage, seed density and initial size of the seeds. For negligible values of the initial coverage of the substrate by seeds, manageable analytical expressions are attained. The model has been validated by comparison with experimental data available in the literature. This study is significant in connection to the possibility of determining optimal growth conditions for ultrathin nanocrystalline diamond (NCD) film. In fact, the knowledge of the seeding/nucleation density that allows a given minimum average thickness of continuous film is of utmost importance for the development of technologically advanced applications.

---

* Corresponding authors.

*E-mail addresses*: tomellini@uniroma2.it (M. Tomellini); polini@uniroma2.it (R. Polini)

# 1  Introduction

Due to the unique combination of outstanding and tunable properties, ultrathin (≤ 100 nm) diamond films have numerous advanced technology applications including micro/nanoelectrochemical systems (MEMS/NEMS), protective optical coatings of sensors and lenses, biomedical coatings, transparent coatings as potential substitutes for indium-tin oxide [ITO], and photon induced electron emitters in solar energy conversion devices [1, 2]. Ultrathin nanocrystalline diamond (NCD) films as thin as 10 nm could allow on/off switching of silicon vacancy (SiV) color center photoluminescence by changing their surface termination [3].

To get to the point, there is a growing interest in the research field of ultrathin diamond coatings deposited by Chemical Vapor Deposition (CVD) on non-diamond substrates.

However, it is a well-established fact that, *i)* due to the high surface energy of diamond, its films form on heterosubstrates via Volmer-Weber, also known as island growth, mechanism, independently of the specific CVD process; *ii)* diamond nucleation from the vapor phase on heterosubstrates is characterized by quite low surface densities, typically $10^4$–$10^6$ nuclei/cm², i.e. about 10 orders of magnitude lower than the typical density of atoms at solid surfaces. Consequently, continuous films, i.e., closed-films where full surface coverage by growing nuclei occurs, having < 100 nm average thickness, or height, can only be grown in the presence of sufficiently large (> $10^{10}$ cm$^{-2}$) nucleation densities. Many techniques have been developed to increase diamond nucleation density. These techniques have been thoroughly reviewed in the literature [1-5]. The methods developed over the last three decades allowed decreasing the thickness of continuous CVD diamond films from 1 μm down to less than 10 nm [1-3]. The most effective technique to date is the so-called "seeding", which relies on the deposition of detonation nanodiamonds (DNDs) on the substrate surface with as high density (cm$^{-2}$) and homogeneity as possible. When the substrate is sonicated in a suspension of deagglomerated DNDs with average size around 3-5 nm, an electrostatic interaction between DNDs and the substrate surface occurs, leaving seed densities as large as $10^{12}$ cm$^{-2}$ [6]. Recently, S. Stehlik *et al* reported seed densities exceeding $10^{13}$ cm$^{-2}$ by using 2 nm seeds [3]. These tiny particles left at the substrate surface cannot be considered as *nucleation sites*, in that they act as diamond "growth centers" when exposed to the activated gas mixture in the CVD reactor. Moreover, with very large surface densities, the surface area fraction covered by seeds could not be neglected (being $\pi N_0 r_0^2$, where $N_0$ is the seeding density and $r_0$ the average seed radius); this has significant implication on the assumption of their random distribution at the substrate surface, as will be discussed below in detail. In any case, it is intuitive that the larger the surface density of growth centers (or *nucleation sites*) during CVD, the thinner will be the continuous diamond film.

Several attempts have been made to predict film coalescence from seeds during CVD. M. Lions *et al* [7] modeled the substrate surface coverage as a function of thickness using an approach similar to that derived by J. B. Austin and R. L. Rickett for the phase transformation of austenite [8] and cited in a milestone paper by Melvin Avrami [9]. By fitting numerical simulations, Lions and coworkers could estimate a critical thickness of 124 nm for a nucleation density of $(3.5 \pm 0.5) \times 10^{10}$ cm$^{-2}$, assuming as 99.0% the full coverage of the substrate surface coated by diamond [7].

Recently, the evolution of seeds during growth, with the formation of closed and porous films, has been studied by S. D. Janssens *et al* [10] in 2D, by using vector-based algorithms developed combining computational geometry and network theory. A detailed study on the microstructure of the 2D-deposit at the film closure transition and connected-grain transition has been carried out depending on initial size of the seeds.

The aim of our paper is to provide an analytical tool for estimating the mean film thickness we should expect for the complete coverage of the substrate surface as a function of seeding density. The role played by the spatial distribution of seeds, linked to their finite size, on the mean thickness of the film will be elucidated and represent one of the goals of the present work. Moreover, the comparison of experimental findings with our analytical model could give information about the stability of seeds under CVD conditions, namely if "all" seeds survived in the early stages of CVD, in that some of them could disappear due to either solid state diffusion of carbon in the substrate or etching/gasification induced by high concentration of atomic hydrogen in the CVD environment [3, 11, 12]. The present modeling therefore provides numerical outputs that can be useful to experimentalists to optimize film growth conditions.

The paper is divided as follows. In section 2.1 we tackle the problem of modeling mean film thickness for growth by spatially correlated seeds, equal in size, in the framework of a mean field approach, where the exact solution of the kinetics is obtained in closed form. Section 2.2 is devoted to the random approximation for the distribution of seeds at the substrate surface. Under these circumstances the approach allows to study the effect of a size distribution of the seed population on mean thickness. Numerical results of the theoretical approach are presented and discussed in section 2.3 and the experimental validation of the model in section 2.4. In the "Conclusion" section we summarize the main results of the work.

## 2 Results and discussion

*2.1 "$2\frac{1}{2}D$ diamond growth" model by nano-diamond seeding*

The growth mode here discussed could be better called $2\frac{1}{2}D$ growth rather than 3D growth, since it describes the growth on the substrate plane (2D space) and along the surface normal (1/2D space, i.e., one-half of the third dimension) of the nanoparticles dispersed on the substate surface [13]. In what follows, nano-seeds are assumed to be hemispheroids, their shape ("extended shape", see below) is conserved during the growth, and the growth law is assumed to be known. The mechanism of growth is ruled by impingement where, upon collision between two seeds, the growth ceases at the common interface. Hemispheroid (prolate or oblate) is one of the simplest shape one can assume in order to deal with anisotropic growth of crystallite [14]. The ratio between the two semiaxes is indicative of the aspect ratio of the nucleus, that is dictated by different growth rates on- and normal to- substrate plane. To be specific, we define $r_0$ as the radius of the projection of the seed on substrate surface, considered circular. Moreover, being in general the height of the seed different from $r_0$, we approximate its shape with a hemispheroid. For a spherical seed this implies assuming an initial aspect ratio[1] of $\alpha = 1$, with the seed growing as a hemispheroid with the same $\alpha$. As far as the seed distribution on the surface is concerned, it is strongly dependent on both their initial size, $r_0$, and surface density, $N_0$. The distribution of seeds can be considered random provided the fraction of substrate surface covered by seeds, $S_0$, is negligible: $S_0 \ll 1$ (see below). For a Dirac delta distribution of the seed-size we get $f(r_0) = \delta(r - r_0)$ and $S_0 = \pi N_0 r_0^2$. Strictly speaking, a random distribution is only attained for $r_0 = 0$ where the evolution of the substrate coverage by diamond can be modeled by the KJMA (Kolmogorov-Johnson-Mehl-Avrami) theory in 2D [9, 15-17]. The reason why for $r_0 \neq 0$ the distribution of seeds is not random is due to the fact that a seed can only lay in the uncovered portion of the substrate surface i.e., in a region not already occupied by any seed. In other words, since two seeds cannot be placed at relative distance shorter than $2r_0$, the seeds are spatially correlated. At $t = 0$ the distribution of seeds on the surface can be envisaged as resulting from a RSA (Random Sequential Adsorption) process, that is a well-known non-Poissonian process [18, and references therein].

The problem of finding the kinetics of $2\frac{1}{2}D$ growth has been firstly faced by Fletcher and Matthews [13] and solved by Bosco and Rangarajan [14], for random nucleation and parabolic growth of nuclei, by exploiting the KJMA model. However, the 2D growth model discussed in these works dealt with progressive nucleation taking place on substrate surface, randomly; therefore, it cannot be applied to the seeding process considered here. To the best of authors knowledge, modeling of $2\frac{1}{2}D$ growth mode in the case of non-random distribution of seeds has not been proposed in literature so far. On the contrary, the kinetics of phase transformations in homogeneous systems by correlated nucleation and growth has been the subject of several studies [19-25]. In these approaches the initial size of nuclei is taken equal to zero as in KJMA approach.

---
[1] The aspect ratio is usually defined as the ratio between the width and the height of nucleus.

In this section, we model the $2\frac{1}{2}D$ growth mode in the case of nano-seeding of the surface, where $r_0 \neq 0$ and – consequently – the distribution is not random. The aim of the present computation is to determine the temporal evolution of the mean thickness of the film as a function of $r_0$ and $N_0$ parameters that characterize the system, and, ultimately, the minimum average thickness (sometimes referred to as "critical thickness" in the relevant literature) of the continuous film deposited by growing $N_0$ seeds per unit area.

We recall that the mean thickness of the film is computed from the total volume of the film, $V'$, and the fraction of the substrate surface covered by the film, $S'$. By denoting with $z(x,y)$ the height of the deposit at point $(x,y)$ on substrate plane, at time $t$ we get: $V'(t) = \int_{D(t)} z(x,y) dx dy$ where the domain of integration is the portion of the substrate surface covered by the film. The measure of the domain is the area covered by the film: $|D(t)| = S'(t)$. The mean height of the film is defined as $\bar{h}(t) = \frac{1}{|D(t)|} \int_{D(t)} z(x,y) dx dy$, that is $\bar{h}(t) = \frac{V'(t)}{S'(t)}$ or, $\bar{h}(t) = \frac{V(t)}{S(t)}$ where $V$ is the film volume per unit of substrate area and $S$ the fraction of substrate surface covered by the deposit.

We consider the case of hemispheroidal seeds, prolate or oblate, with semi axes $r$ and $h_{max}$, respectively, on- and normal to- the substrate plane. $h_{max}$ is the maximum height of the growing seed (see Fig. 1). As anticipated, at $t = 0$ we consider a Dirac delta distribution of seeds all with the same size and surface density $N_0$. At the beginning of the growth the fractional coverage equals $S_0 = \pi N_0 r_0^2$. We define the growth law for the seed substrate interface as $r(t) = r_0 + \int_0^t G(\tau) d\tau$, $G(\tau)$ being the growth rate. Also, the aspect ratio is taken as constant during the growth, namely $\eta = \frac{h_{max}(t)}{r(t)} = $ constant . We point out that for the growth rate of well separated diamond particles, $G(\tau) = constant$ has been reported which implies interfacial limited growth at the nucleus-gas interface [26-28].

For hemispheroidal nuclei with constant aspect ratio $\left(\alpha = \frac{2}{\eta}\right)$, the volume and lateral surface of the nucleus are $v = \frac{2}{3}\pi \eta r^3$ and $s = 2\pi r^2 w(\eta)$, respectively, with $w(\eta)$ a function of $\eta$. For interfacial limited growth at gas/nucleus interface we get $\frac{dv}{dt} \propto s$, which implies $\frac{dr}{dt} = G$ and $\frac{dh_{max}}{dt} = \eta G$, i.e., constant growth rates of semi axes (depending on $\eta$).

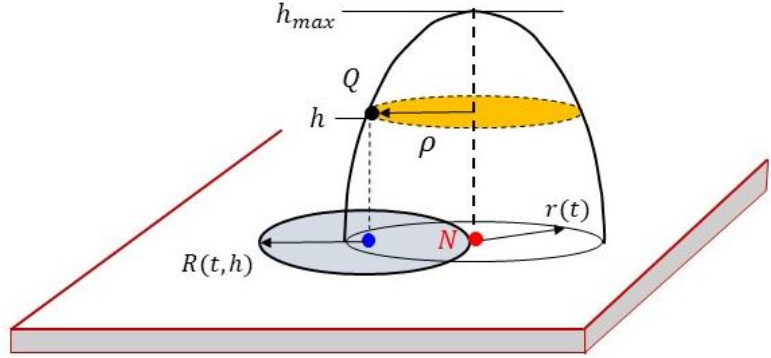

**Fig. 1** Schematic representation of a hemispheroidal seed on a solid surface at running time $t$, where $r(t)$ is the radius of its circular base. The generic point of the space, $Q$, located at height $h$, is not transformed at time $t$ provided no seed falls within the blue disk of radius $R^2 = \rho^2 = r^2(1 - \frac{h^2}{h_{max}^2})$, where $(\rho, h)$ are the coordinates of a point of the hemispheroid surface. In fact, seeds located within such a disk are capable of transforming $Q$ before time $t$. The stochastic problem is therefore equivalent to a stochastic process of dots in 2D-space. The hemispheroid centered at $N$ represents a seed just at the border of the blue disk. $h_{max} \equiv h_{max}(t)$ is the maximum height of the nucleus at time $t$. For the hemispheroid the aspect ratio is equal to $\alpha = 2r/h_{max}$.

The computation of the mean thickness requires the knowledge of $V$ and $S$ that are performed based on the probability theory. In particular, the key quantity of the approach is the so called "exclusion (or void) probability", $P_0(R)$, that is the probability that, in a D-dimensional space (in our case 2D space, i.e., the substrate plane), no center of seed is found within a region of volume $v_D(R)$ (in our case $v_2(R) = \pi R^2$). Throughout the paper we refer to $R$ as the size of the "exclusion zone". Were the distribution of seeds random, we would have

$$P_0(R) = e^{-N_0 \pi R^2} \tag{1}$$

in agreement with the Poisson process. Because $r_0 \neq 0$ the seeds are spatially correlated (on the substrate plane) according to a hard-disk potential.

The statistical mechanics of many particles system with hard-core interaction is an important topic in connection with the study of condensed matter in the liquid state. In particular, for hard-core potential the lowest order term of the radial distribution function is $g(r) \cong H(r - \sigma)$, where $H(x)$ is the Heaviside step-function[2] and $\sigma$ the minimum relative distance accessible to a couple of particles [29]. To the purpose of the present computation, we refer to the results by Torquato *et al.* which provide exact series representation for the probability functions of systems of spheres which interact with arbitrary potential [29, and references therein]. The coefficient of the series are multidimensional integrals involving *n*-particles probability density functions, $\rho_n(\mathbf{r}_1, \ldots, \mathbf{r}_n)$, where $\rho_n d\mathbf{r}_1 \ldots d\mathbf{r}_n$ is the probability of finding *n* particles in volume elements $d\mathbf{r}_1, \ldots d\mathbf{r}_n$, at the stated positions, independently of the position of the other *N-n* particles. The method gives the expression for the void and the nearest particle probability functions for a system of hard-sphere by exploiting the scaled-particle theory [31]. To be specific, the void probability is defined as the probability of finding a region, which is a D-dimensional spherical cavity of radius $R$, empty of particle centers [30]. For the present study, we made use of the void probability for an ensemble of hard disk of radius $r_0 = \frac{\sigma}{2}$, $\sigma$ being the minimum distance between two particles, in our system seeds on the surface. In terms of the quantities above defined, $S_0$ and $r_0$, this probability is given by

$$P_0(R) = (1 - N_0 \pi R^2) H(r_0 - R) + (1 - S_0) e^{-\frac{S_0}{(1-S_0)^2}\left[\left(\frac{R}{r_0}\right)^2 - 2S_0\left(\frac{R}{r_0} - 1\right) - 1\right]} H(R - r_0) \qquad (2)$$

where $H(x)$ is the Heavisde step-function; it is worth recalling that $r_0$ and $S_0$ are the radius of the seeds and the fractional surface coverage at $t = 0$, respectively. It is easy to verify that in the limit $r_0 \to 0$ (for constant $N_0$ in the $S_0 = \pi N_0 r_0^2$ expression) eqn. 2 leads to eqn. 1 that holds for random distribution of seeds.

To model the temporal evolution of the volume of the deposited material, let us consider a generic point of the space, say $Q$, located at height $h$ from the surface, and determine the probability that the point $Q$ is not transformed by the new phase at time $t$ (Fig.1). This event requires that (at $t = 0$) no seed lays at a distance shorter than $R = R(t, h)$ from the projection of point $Q$ on the substrate, being $r = r(t)$ the radius of the base of the seed at time $t$ (see Fig.1). This probability is therefore equal to the exclusion probability (eqn. 2) with the size of the exclusion zone $R(t, h)$. To determine $R(t, h)$ we employ the equation of the ellipsoid, $\frac{\rho^2}{r^2} + \frac{h^2}{h_{max}^2} = 1$, for the coordinate $(\rho, h)$ of a point of the surface (Fig. 1). Since $\rho(h)^2 = r^2(1 - \frac{h^2}{h_{max}^2})$ we get $R(t, h) = \rho(h, t) = \sqrt{r(t)^2 - \frac{h^2}{\eta^2}}$ that is $R^2(t, h) = (r_0 + \gamma(t))^2 - \frac{h^2}{\eta^2}$, where $\gamma(t) = \int_0^t G(\tau) d\tau$. It follows that the probability that the point $Q$ is transformed into the new phase within time $t$ is given by $(1 - P_0)$.

---

[2] The Heaviside function is defined as: $H(x) = 0, x < 0$; $H(x) = 1, x > 0$.

At running time $t$, the volume of the film per unit area is eventually given by integration over $h$:[3]

$$V(t) = \int_0^{h_{max}(t)} (1 - P_0(h,t))dh, \tag{3}$$

where the maximum height is linked to the seed radius by $h_{max} = \eta r(t)$. By using in eqn. 2 the reduced variable $y = \frac{h}{\eta r(t)}$, eqn. 3 becomes,

$$V(t) = \eta r(t) \int_0^1 \left[1 - (1 - N_0 \pi r^2(1-y^2))H\left(y^2 + \left(\frac{r_0}{r(t)}\right)^2 - 1\right)\right.$$
$$\left. - (1-S_0)e^{-A(y)} H\left(1 - y^2 - \left(\frac{r_0}{r(t)}\right)^2\right)\right] dy, \tag{4a}$$

with

$$A(y) = \frac{S_0}{(1-S_0)^2}\left[\left(\frac{r(t)}{r_0}\right)^2 (1-y^2) - 2S_0\left(\frac{r(t)}{r_0}\sqrt{1-y^2} - 1\right) - 1\right].$$

By defining the time dependent variable $z = \frac{r(t)}{r_0} \geq 1$,[4] and considering the contributions of the Heaviside functions, one gets

$$\bar{h}(t) = \frac{V(t)}{S(t)} = \frac{\eta r(t)}{S(t)}\left(1 - \int_0^{\sqrt{1-\frac{1}{z^2}}} (1-S_0)e^{S_0} e^{-\frac{S_0 z^2}{(1-S_0)^2}\left(\sqrt{1-y^2} - \frac{S_0}{z}\right)^2} dy\right.$$
$$\left. - \int_{\sqrt{1-\frac{1}{z^2}}}^1 (1 - S_0 z^2(1-y^2))dy\right). \tag{4b}$$

In eqn. 4b the time dependence of $V(t)$ is due to the $r(t)$ function which also enters the $z(t)$ term. Concerning the fraction of substrate surface covered by the film, it is given by

$$S(t) = 1 - P_0(0,t), \tag{5a}$$

namely,

$$S(z) = 1 - (1-S_0)e^{-\frac{S_0}{(1-S_0)^2}[z^2 - 2S_0(z-1) - 1]}. \tag{5b}$$

---

[3] The term $1 - P_0$ is equal to $a(h)$ that is the ratio between the area of the section of the film at $h$ and the whole substrate area. The differential volume of the film, per unit area, is $dV(h) = a(h)dh$.
[4] To simplify the notation in some occurrences we omit the time dependence in z.

In eqn. 5$b$, $S$ is a function of time through the $z(t)$ function defined above. It is also useful to express the kinetics in terms of the $S$ variable rather than of $t$. To this end, we express $r$ as a function of $z$ according to $r(S) = z(S)\sqrt{\frac{S_0}{\pi N_0}}$, where,

$$z(S) = S_0 + \frac{1-S_0}{\sqrt{S_0}}\sqrt{S_0 - \ln\frac{1-S}{1-S_0}} \qquad (5c)$$

as obtained through inversion of eqn. 5$b$. The average thickness, as a function of fractional coverage, $S$, is eventually given by

$$\bar{h}(t) = \eta \frac{z(S)}{S}\sqrt{\frac{S_0}{\pi N_0}}\left(1 - \int_0^{\sqrt{1-\frac{1}{z^2}}}(1-S_0)e^{S_0}e^{-\frac{S_0 z^2}{(1-S_0)^2}\left(\sqrt{1-y^2}-\frac{S_0}{z}\right)^2}dy - \Phi(S)\right), \qquad (6a)$$

where

$$\Phi(S) = \frac{1}{3}\left[\frac{\sqrt{z^2-1}}{z}(2S_0 z^2 + S_0 - 3) + (3 - 2S_0 z^2)\right] \qquad (6b)$$

and $z = z(S)$ is given by eqn. 5$c$.

The present theoretical approach (eqns. 6), indicates the following scaling of the mean thickness with surface coverage of the deposit:

$$\bar{h} = N_0^{-\frac{1}{2}}g(S, S_0, \eta) \qquad (7)$$

where $g(S, S_0, \eta)$ depends upon seed shape and initial surface coverage, that is by the correlation degree of the seed ensemble. The behavior of the closed form solutions (eqns. 6), as a function of density, surface coverage and correlation degree among seeds, will be discussed in § 2.3.

*2.2 The random approximation.*

The random approximation implies using the void probability, given by eqn. 1, in eqn. 3 where the surface coverage is now given by the KJMA theory: $S(t) = 1 - e^{-N_0 \pi r(t)^2}$. Strictly speaking, as a necessary condition the random distribution can only be realized with point-like seeds.

To begin with, we develop the modeling for the growth of 0D seeds ($r_0 = 0$) and, successively, for a distribution of seeds not equal in size. For $r_0 = 0$ and hemispheroidal seeds the mean thickness of the deposited material is given by

$$\bar{h}(t) = \frac{V(t)}{S(t)} = \frac{1}{S(t)}\int_0^{\eta r}(1 - e^{-N_0\pi(r^2 - \frac{h^2}{\eta^2})})dh = \eta\frac{r(t)}{S(t)}\int_0^1(1 - e^{-S_x(1-y^2)})dy \qquad (8a)$$

where $S_x = S_x(t) = N_0\pi r(t)^2$ is the "extended surface" defined in the KJMA theory [16]. Eqn. 8a can be rewritten as a function of surface coverage as

$$\bar{h}(S) = \eta\frac{1}{\sqrt{\pi N_0}}\frac{\sqrt{S_x} - F_D(\sqrt{S_x})}{1 - e^{-S_x}}, \qquad (8b)$$

where $S_x = S_x(S)$ is given by

$$S_x(S) = -\ln(1 - S) \qquad (8c)$$

and $F_D(x)$ is Dawson's integral, defined as $F_D(x) = e^{-x^2}\int_0^x e^{z^2}dz$. For a random distribution of seeds, the present approach can be improved to deal with seed-size distribution functions. Let us group the population of seeds in classes of seeds, equal in size, and denote with $r_{0,i}$ and $N_i$ their radius and number density, respectively. To determine the volume of the deposit, we employ the probabilistic approach outlined above, where the "void probability" of each class is given by eqn. 1. The whole "void probability", that enters eqn. 3, is attained by considering the independent events in which the hemispheroidal seeds of the $i^{th}$ class do not lay in the disk of radius $R_i = \sqrt{r_i(t)^2 - \left(\frac{h}{\eta}\right)^2} = \sqrt{[r_{0,i} + \gamma(t)]^2 - \left(\frac{h}{\eta}\right)^2}$ where the aspect ratio is assumed to be independent of seeds class. The void probability therefore reads

$$P_0(h, t) = \prod_i e^{-N_{0,i}\pi R_i^2} = e^{-\sum_i N_{0,i}\pi R_i^2}. \qquad (9a)$$

To perform the continuum limit of eqn.9a we define the size distribution function of seeds, $f(x)$, where $f(x)dx$ is the probability that the initial size of the seed lays between $x$ and $x + dx$. This probability density function (PDF) is normalized according to $\int_0^\infty f(x)dx = 1$. In the continuum limit the argument of the exponential in eqn. 9a and the extended surface become, respectively,

$$\sum_i N_{0,i}\pi R_i^2 = N_0\pi\int_0^\infty f(x)\left[(x + \gamma(t))^2 - \left(\frac{h}{\eta}\right)^2\right]H\left(x + \gamma(t) - \frac{h}{\eta}\right)dx, \qquad (9b)$$

$$\sum_i N_{0,i}\pi r_i^2 = N_0 \pi \int_0^\infty f(x)(x+\gamma(t))^2 dx, \tag{9c}$$

where $N_0$ is the total density of seeds. The surface coverage is given by eqn. 5a, namely $S(t) = 1 - e^{-S_x(t)}$, with the extended surface (eqn. 9c)

$$S_x(t) = N_0\pi[\langle x^2\rangle + 2\gamma(t)\langle x\rangle + (\gamma(t))^2], \tag{10}$$

where $\langle\cdot\rangle$ denotes mean values over the $f(x)$ probability density function. The mean thickness of the film is finally given by

$$\bar{h}(t) = \frac{\eta}{1-e^{-S_x}}\left[\int_0^{\gamma(t)}\left(1 - e^{-[S_x - N_0\pi\xi^2]}\right)d\xi + \int_{\gamma(t)}^\infty (1 - e^{-\varphi(\xi)})\,d\xi\right], \tag{11a}$$

where

$$\varphi(\xi) = \pi N_0 \int_{\xi-\gamma(t)}^\infty f(x)[(x+\gamma(t))^2 - \xi^2]dx. \tag{11b}$$

The last integral in eqn.11a does not diverge since, at large $\xi$, $\varphi(\xi)$ is expected to decrease, rapidly, being $S_0 = N_0\pi\langle x^2\rangle < 1$.

We applied eqns. 11a, 11b to describe the growth of seeds, equal in size, i.e., with the PDF $f(x) = \delta(x - r_0)$. It is possible to show that in the $S$ domain, the mean thickness is given by eqn. 8a, independently of the $r_0$ value. Under these circumstances the scaling of $\bar{h}$ is given by eqn. 7 for $S_0 \to 0$.

*2.3 Numerical computations.*

In this section we report the typical behaviors of the main quantities of the model useful for characterizing the evolution of film formation. Fig. 2A shows the evolution the fraction of substrate surface covered by the deposit as a function of mean thickness for spatially correlated hemispherical seeds (eqns. 6a, 6b). The curves have been computed for several values of correlation degree of the seed population, namely $S_0 = \pi N_0 r_0^2$, and at $N_0 = 400$ μm$^{-2}$ ($4\times 10^{10}$ cm$^{-2}$). The computations of Fig. 2 are done for $S_0$ values lower than that at the jamming point, for RSA in 2D space, namely $S_0 \cong 0.542$ [32] or $S_0 \cong 0.547$ [33, 34]. The initial value of $S$ is $S_0$, for which the mean thickness is equal to $\bar{h}(0) = \frac{2r_0}{3} = \frac{2}{3}\sqrt{\frac{S_0}{\pi N_0}}$.

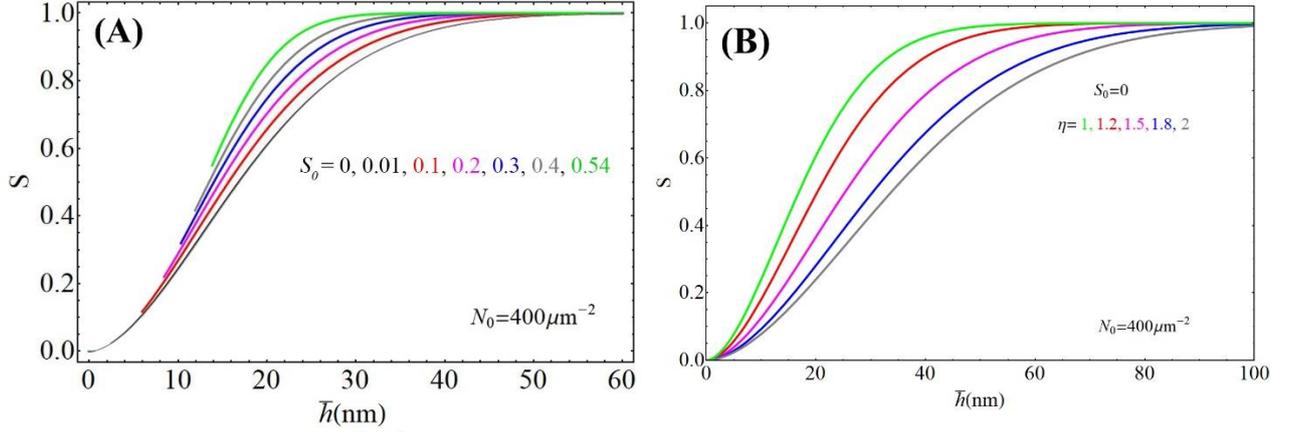

**Fig. 2** Fraction of substrate surface covered by a deposit with spatially correlated hemispherical seeds ($\eta = 1$) providing an initial $S_0$ coverage (panel A). Surface density of seeds is set equal to 400 µm$^{-2}$ ($4\times10^{10}$ cm$^{-2}$). The plots are calculated by using eqns. 6a, 6b and 8a, 8b ($S_0 = 0$). In panel B the $S(\bar{h})$ function is displayed for several values of the parameter η (prolate spheroid) for random distribution of seeds.

The random case is expected to be recovered for $S_0 \ll 1$. In fact, the black curve, at $S_0 = 10^{-2}$, is nearly indistinguishable from that for uncorrelated seeds ($S_0 = 0$, curve in gray); the difference between the two curves is of the order of magnitude of $10^{-3}$. It stems that for this case seed distribution is random-like. Now, assuming $r_0 \approx 3$ nm (a typical value for non-agglomerated detonation nanodiamond, DND, seeds) and $N_0 = 400$ µm$^{-2}$, we get $S_0 \approx 0.01$. From the plot of Fig. 2 we can estimate a critical thickness of around 50 nm (black curve).

It is worth noting that in our analytical modeling, which works in the continuum limit and for an infinite system (thermodynamic limit), the value $S = 1$ is reached asymptotically. This is a typical behavior linked to this kind of approaches, as pointed out in literature [7, 10]. In view of possible use of the present model by experimentalists, it is therefore necessary to define a final coverage value ($S_f$) which is representative of the minimum thickness at film closure. For instance, in ref. [7] a value of $S_f = 0.99$ has been set in connection with author's modelling. Here we set the $S_f$ value also in connection with the degree of smoothness of the film. In fact, the main goal of the seeding technique is the attainment of thin, continuous, and smooth films. On this basis, we define a sort of "smoothness" degree of the film as $\Delta v = \frac{v_s - v}{v} = \frac{h_{max} - \bar{h}}{\bar{h}}$, where $v_s$ is the volume of a completely smooth deposit with maximal thickness $h_{max}$. By choosing $S_f$ in such a way that $\Delta v \leq 10\%$ in the significant range of $S_0$, we get $S_f = 0.999$ in the theoretical curves.

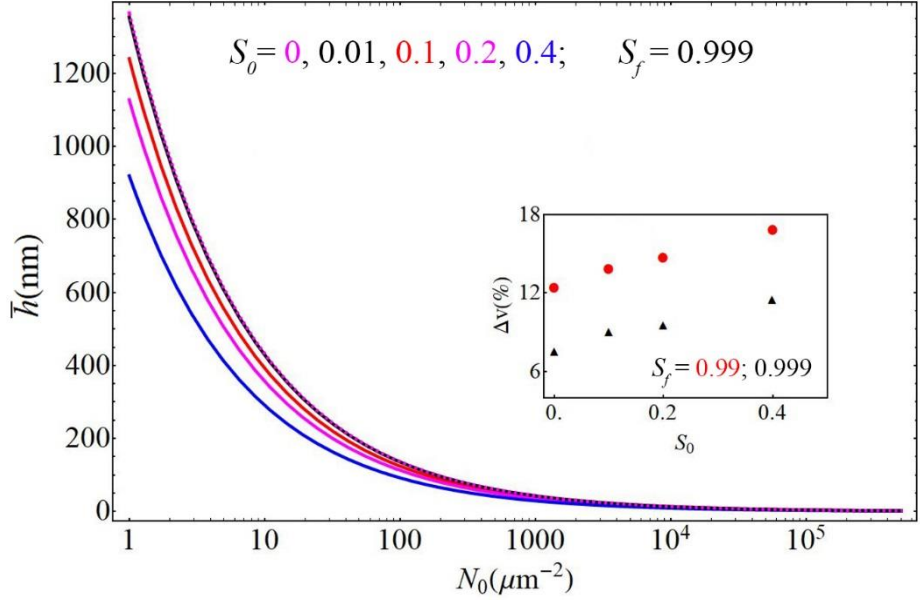

**Fig. 3** Behavior of the mean thickness as a function of seed density and correlation degree ($S_0$) at film closure ($S_f = 0.999$). Computations refer to hemispherical nuclei. The curve at $S_0 = 0$ (dashed line) is almost indistinguishable from that at $S_0 = 0.01$ (solid black line). The $S_f$ value is chosen in order to get a quite smooth deposit with $\Delta v \leq 10\%$ (see text). In the inset $\Delta v$ is reported as a function of $S_0$ for two values of coverages (0.99 and 0.999).

Based on this choice, in Fig. 3 we plot the function, eqn.7 at $\eta = 1$, by computing the $g$ function at $S_f = 0.999$ and for various values of $S_0$. It follows that at given $N_0$ the mean thickness decreases with $S_0$. In accord with the results for the $S(\bar{h})$ function (Fig. 2A), the scaling of $\bar{h}$ with $N_0$ at $S_0 = 0.01$ coincides with the random case (dashed line, Fig. 3). In Tab.1 we summarize the results of the proposed model in a scheme that can be worthwhile in connection with experimental studies on critical thickness. Specifically, we report the range of variation of mean thickness at $S_f$ in the interval of aspect ratios $1 < \alpha < 1.8$ for several values of $S_0$ and seed surface density. In Tab.2 we report the initial radius of the seed for the various values of $N_0$ and the initial fractional coverages of Tab.1.

*2.4 Experimental validation of the model.*

Scanning Electron Microscopy (SEM) and Atomic Force Microscopy (AFM) can be used to measure the minimum average thickness at which full surface coverage occurs and to experimentally validate the model. However, few reliable data are available in the literature. Fig. 4 displays the log-log plot of the calculated mean thickness as a function of seed density values exceeding 100 µm$^{-2}$ ($N_0 > 1 \times 10^{10}$ cm$^{-2}$), i.e., a surface density interval necessary for obtaining continuous nanocrystalline diamond (NCD) films. In the same plot, data from literature have been reported. We point out that in the considered papers the $S_0$ values are not provided, apart from the study of ref. [3].

| $N_0/\mu m^{-2}$ | $S_0 = 0$ $\bar{h}(\eta = 1.1 \div 2)/nm$ | $S_0 = 0.1$ $\bar{h}(\eta = 1.1 \div 2)/nm$ | $S_0 = 0.15$ $\bar{h}(\eta = 1.1 \div 2)/nm$ | $S_0 = 0.2$ $\bar{h}(\eta = 1.1 \div 2)/nm$ |
|---|---|---|---|---|
| 200 | 106-193 | 97-176 | 92-168 | 88-160 |
| 400 | 75-136 | 68-124 | 65-119 | 62-113 |
| 600 | 61-112 | 56-101 | 53-97 | 51-93 |
| 800 | 53-97 | 48-88 | 46-84 | 44-80 |
| 1000 | 48-86 | 43-79 | 41-75 | 39-72 |
| 2500 | 30-55 | 27-50 | 26-48 | 25-45 |
| 5000 | 21-39 | 19-35 | 18-34 | 18-32 |
| 7500 | 17-32 | 16-29 | 15-27 | 14-26 |
| 10000 | 15-26 | 14-25 | 13-24 | 12-23 |

**Table 1** Interval of variation of the mean thickness for $1.1 < \eta < 2$ (i.e., $1 < \alpha < 1.8$) for several values of seed densities and $S_0 = 0, 0.1, 0.15$ and $0.2$ ($S_f = 0.999$). Mean thicknesses values are approximated by integers.

| $N_0/\mu m^{-2}$ | $S_0 = 0.1$ $r_0/nm$ | $S_0 = 0.15$ $r_0/nm$ | $S_0 = 0.2$ $r_0/nm$ |
|---|---|---|---|
| 200 | 12.6 | 15 | 17.8 |
| 400 | 8.9 | 11 | 12.6 |
| 600 | 7.3 | 8.9 | 10.3 |
| 800 | 6.3 | 7.7 | 8.9 |
| 1000 | 5.6 | 7 | 7.9 |
| 2500 | 3.6 | 4.4 | 5 |
| 5000 | 2.5 | 3.1 | 3.6 |
| 7500 | 2.1 | 2.5 | 2.9 |
| 10000 | 1.8 | 2.1 | 2.5 |

**Table 2** Seed radius in dependence of surface density, $N_0$, and initial values of fractional coverage, $S_0$.

It's worth recalling here that several literature data concerning the critical thickness might be affected by systematic errors since, either the measured film thicknesses were larger than the critical thickness (e.g., the authors did not stop the deposition at full coverage occurrence), or not all seeds initially dispersed on the substrate surface survived during CVD. The etching of diamond seeds by the H-rich atmosphere employed for the diamond growth from the gas phase has been known since more than two decades [35]. As clearly stated recently by S. Stehlik and coworkers, high temperature (700 °C) and high plasma densities (i.e., high concentrations of monohydrogen in the activated gas mixture) "might lead to a certain reduction of the nucleation density due to possible etching of nanodiamond seeds" [3]. Actually, by using conventional micro-wave CVD (MWCVD), after 8 min deposition, and starting from a seed density of $4\times10^{11}$ cm$^{-2}$, the authors got a non-continuous deposit mainly composed of separated diamond crystals. By counting the number of crystallites deposited after 8 min CVD (Fig. S3(a) of ref. [3]) we have determined a grain density of $1.2\times10^{11}$ cm$^{-2}$, corresponding to 70% loss of initial seeds. Although our estimate could be much approximate, in that based on a single SEM micrograph, it is evident that a conspicuous loss of seeds occurred anyway.

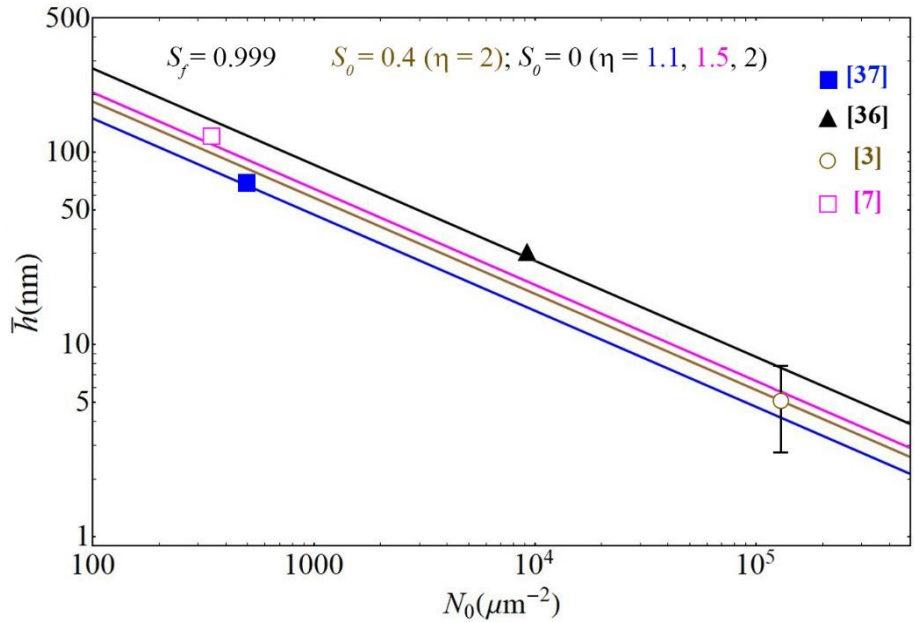

**Fig. 4** Mean thickness of continuous ($S_f$ =0.999) films as a function of seed/nucleation density and both correlation degree ($S_0$) and η parameter, of growing crystallites. Besides the calculated curves, experimental data from refs. [3], [7], [36] and [37] are reported.

To suppress these effects, the authors performed seeds growth at lower temperature (460 °C) and low pressure (0.15 mbar) using a pulsed micro-wave (MW) plasma with a linear antenna arrangement. In this way, most of the 2-nm seed they applied to the silicon substrate with surface density as large as ~ $1.3 \times 10^5$ µm$^{-2}$ ($1.3 \times 10^{13}$ cm$^{-2}$, $S_0 = 0.40$), are supposed to survive during the subsequent 5 h growth step, leading to an ultrathin diamond film of just $5 \pm 2.5$ nm (open circle in Fig. 4). Our analytical model predicts, by using $S_0 = 0.40$, an average thickness of 5.1 nm with $\eta = 2$. Also, for a nearly isotropic seed growth ($\eta = 1.1$), the mean thickness estimated from the present approach ($\approx 4$ nm) is within the error bar of experimental data.

H.-J. Lee and coworkers [36] succeeded in depositing a NCD film of 30 nm starting from an initial seed density of $9.3 \times 10^3$ µm$^{-2}$ (black triangle in Fig. 4). O. Ternyak *et al.* [37] reported the formation of films as thin as 70 nm using a seed density of 500 µm$^{-2}$ (full square in Fig. 4). M. Lions and coworkers [7] calculated the formation of a continuous film of 124 nm from a nucleation density of $(350 \pm 50)$ µm$^{-2}$ (open square in Fig. 4). These experimental data are in good agreement with our model by using aspect ratios, $\alpha = 2/\eta$, in the range 1–1.8 ($1.1 < \eta < 2$) and $S_0 = 0$ that implies a random arrangement of seeds as also hypothesized in [7].

Finally, in Fig. 5 the experimental data of surface coverage as a function of the mean thickness of diamond film, taken from Fig. 3(a) of ref. [7] for $N_0 = 300$ $\mu m^{-2}$, are reported as symbols and described with our model using $S_0 = 0$, $N_0 = 300$ $\mu m^{-2}$, and $\eta = 1.3$.

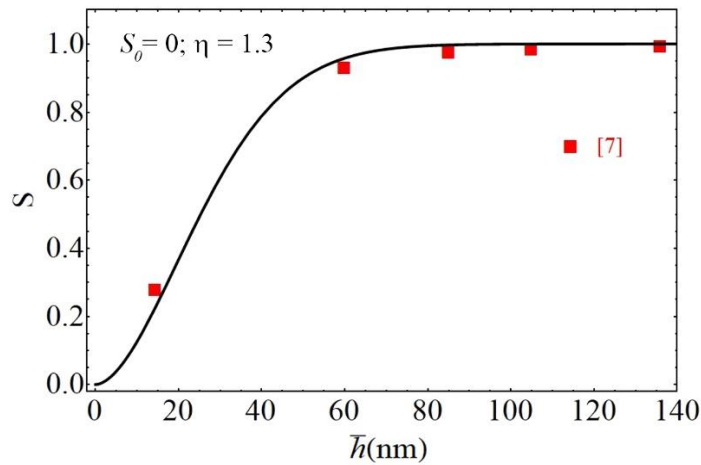

Fig. 5  Trend of the fractional surface coverage as a function of mean thickness of the deposit. Full symbols are the experimental data of ref. 7 for a nucleation density of 300 µm$^{-2}$. Solid line has been computed from eqn. 8*b* in the random approximation ($S_0 \approx 0$) and for $\eta = 1.3$.

## 3   Conclusions

The mean field approach developed in this work highlights the role played by the surface density, size and shape of seeds on critical mean thickness of continuous films. The core of the model is the inclusion of spatial correlation effect among seeds in describing the growth kinetics. It is shown that correlation effects are not negligible (compared with the random case) for $S_0 > 0.1$, therefore becoming relevant, in general, for attaining ultrathin films. For low $S_0$ values the random approximation holds true and manageable analytical expressions are obtained for the mean thickness as a function of surface coverage, seed density and aspect ratio. The surface coverage at film closure is set equal to $S = 0.999$ based on simple considerations on the smoothness of the film. The present computation does show that the mean thickness, and critical thickness as well, as function of seeding density, is in accord with the power law eqn. 7, where $\bar{h} \propto N_0^{-1/2}$. The approach deals with the model case of hemispheroidal seeds, which is the simplest shape that mimics both isotropic ($\eta = 1$) and anisotropic ($\eta \neq 1$) growth. The validity of the analytical approach has been checked through comparison with experimental data available in literature. A good agreement of both critical thickness vs. $N_0$ and surface coverage, $S$, vs. $\bar{h}$, has been obtained in the significant range of aspect ratio $\alpha = 1 - 2$.